\documentstyle[12pt]{article}
\begin{document}

\bigskip

\begin{center}
{\huge {\bf Quantum Creation of }}

{\huge {\bf Closed Universe with Both }}

{\huge {\bf Effects of Tunneling and Well}}

\bigskip \bigskip

{\Large {\bf De-Hai Zhang}}\\[0pt]

\bigskip 

(e-mail: dhzhang@sun.ihep.ac.cn)\\[0pt]

Department of Physics, Graduate School in Beijing,\\[0pt]

University of Science and Technology of China,\\[0pt]

P.O.Box 3908, Beijing 100039, P.R.China.\\[0pt]

\bigskip

{\bf Abstract:}
\end{center}

{\hspace*{5mm}A new ''twice loose shoe'' method in the Wheeler-DeWitt
equation of the universe wave function on the cosmic scale factor }$a${\ and
a scalar field $\phi $} is suggested in this letter.{\ We analysis the both
affects come from the tunneling effect of }$a${\ and the potential well
effect of $\phi $, and obtain the initial values }$a_0$ and $\phi _0${\
about a primary closed universe which is born with the largest probability
in the quantum manner. Our result is able to overcome the ''large field
difficulty'' of the universe quantum creation probability with only
tunneling effect. This new born universe has to suffer a startup of
inflation, and then comes into the usual slow rolling inflation. The
universe with the largest probability maybe has a ''gentle'' inflation or an
eternal chaotic inflation, this depends on a new parameter }$q$ which
describes the tunneling character{.}

\bigskip

\section{Introduction}

{\hspace*{7mm}}A lot of observations have proved that our universe undergoes
a big bang$^{[1]}$ from a high temperature and high density state. Today our
universe is very large and old, very isotropic and homogeneous in the large
scale, and almost flat. In order to illustrate all of these features, our
universe in its very early period needs a more violence expansion, i.e.,
inflation$^{[2]}$ which must has an enough large inflating e-folds and a
small background fluctuation. But somebody would like to ask what is before
the inflation? Why did the universe inflate? Further he could ask whether
the universe had its birth? How was the universe born? These are very
interesting problems the wise mankind want to probe into. Some scientists
have supplied for us a part of possible answers about these questions by
right of their insights. The universe may be born in quantum way$^{[3]}$,
and it may suffer eternal chaotic inflation$^{[4]}$. Maybe it is born from
an instanton$^{[5]}$. However, the present theories have still some
difficulties, many details are still unclear.

We all hope that a theory will be simple as possible as. However for a most
typical potential, such as $m^2\phi ^2/2$, in chaotic inflation theory, it
meets some difficulty which can be seen by following analysis. At first
there is a famous probability formula about the quantum birth of the
universe, $\rho \propto \exp [-24\pi ^2M_p^4V^{-1}(\phi )]$, in the quantum
cosmology$^{[6]}$, where $8\pi G=M_p^{-2}=\kappa ^2$, and $M_p=2.4\times
10^{18}$GeV is the Planck mass. If this formula is earnest, the largest
probability happens at a place where the field value $\phi $ is largest,
then maybe it tends to infinity. However if $\phi $ exceeds over a critical
value, the universe will undergo an excessive inflation and never roll down.
Or we choose a special potential, or we need a cutting off or a suppression
of the probability at large field limit. We call it as ''large field
difficulty''.

In this paper we find out that a potential well effect appeared naturally in
the quantum cosmology can overcome this difficulty. My starting point is
still the famous Wheeler-DeWitt equation$^{[7]}$ of the universe wave
function on the cosmic scale factor $a$ and a scalar field $\phi $. As a
difference from the former investigations, I study both quantum behaviors
about $a$ and $\phi $, i.e., a tunneling effect for $a$ and a potential well
effect for $\phi $ (the latter is new for me). Just this well effect supply
a probability suppression of large field limit. Just these two combined
effects of the tunneling and the well determine that a primary universe has
to have a most possible initial state, which initial values may be suitable
to develop a relevance universe for us through the startup of its inflation.
If this initial field is not too large, the universe maybe doesn't have an
eternal chaotic inflation, therefore this probability makes sense.

In order to illustrate my idea and results, at first we distinguish three
inflationary states in the section 2. Then we review the tunneling effect
and show ''large field difficulty'' in the section 3. We study the well
effect and introduce an important parameter $q$ to describe the tunneling
character in the section 4. The startup, persistent and ending of inflation
are studied in the section 5. We give a more discussion about our results in
the section 6. Finally we point out its progress and shortcoming of our
method in the last section.

\section{Gentle, chaotic and excessive inflations}

{\hspace*{7mm}}Among various scenarios to describe the birth and evolution
of the universe, the eternal chaotic inflation is an important one. The
necessary condition at which the eternal chaotic inflation happens is that
the initial field value $\phi _0$ of inflaton is near a critical one $\phi _c
$ (index $c$ means critical). This critical value is determined by the
condition that the average quantum fluctuation $\Delta \phi _{qu}$ is near
the classical rolling down value $\Delta \phi _{cl}$ of field in a Hubble
time. In the case of potential $m^2\phi ^2/2$, which is regarded as a
typical one for chaotic inflation and is studied mainly in this paper, this
critical value is$^{[8]}$ 
\begin{equation}
\label{crit}\phi _c=2\cdot 6^{1/4}\pi ^{1/2}\kappa ^{-3/2}m^{-1/2},
\end{equation}
which is depend on the mass parameter heavily.

Let us estimate rough mass at first. The constraint on mass comes from the
small fluctuation about $10^{-5}$ of the Cosmic Microwave Background
Radiation (CMBR)$^{[9]}$,%
$$
\delta ^2=150^{-1}\pi ^{-2}\epsilon _e^{-1}\kappa ^4V_e=10^{-10}, 
$$
where $\epsilon _e\simeq 1$ (index $e$ means ending of inflation) is the
slow rolling parameter $\epsilon =\kappa ^{-2}V^{\prime 2}V^{-2}/2$ of the
ending point of inflation. From it we obtain the mass parameter $m\simeq
4\times 10^{-4}\kappa ^{-1}\simeq 10^{15}$GeV or little small, which is just
about the grand unification energy scale. Then the field critical value of
chaotic inflation is $\phi _c=280\kappa ^{-1}$ and the potential is $%
V_c=6\times 10^{-3}\kappa ^{-4}$. We think that up to $V_q\simeq \kappa ^{-4}
$, i.e., $\phi _q\simeq 3500\kappa ^{-1}$ (index $q$ means quantum), the
classical gravity maybe should be applicable.

The eternal chaotic inflation is not the unique possible case. When $\phi
_0\ll \phi _c$, it is obvious that the eternal chaotic inflation can not
happen due to the quantum fluctuation too small, but it has a normal
inflation if $\phi _0\geq \phi _e$ where $\phi _e$ is the field value at the
ending of inflation, which is determined by the end of slow rolling $%
\epsilon _e\simeq 1$ and gets $\phi _e=\sqrt{2}\kappa ^{-1}$ for potential $%
V=m^2\phi ^2/2$. We call it as ''gentle'' or ''classical'' inflation. When $%
\phi _0\gg \phi _c$, a tiny classical rolling down is submerged in a huge
quantum fluctuation of $\phi $, then in spite of the universe undergoes an
eternal chaotic inflation, but it can roll down never and nowhere, any slow
rolling down universe is not produced in its fraction. Maybe the field value
can climb higher and higher. We call this case as ''excessive'' or
''quantum'' inflation. Of course we only call the case $\phi _0\approx \phi
_c$ (maybe its range is rather wide) as standard ''chaotic'' or ''critical''
inflation. Therefore the chaotic inflation which can produce an observed
universe is conditional.

\section{The tunneling effect of scale factor}

{\hspace*{7mm}}The birth of the universe is a complicated problem. If we
predigest this problem to freeze out all of freedoms except the cosmic scale
factor $a$ and a scalar field $\phi $, we can obtain the famous
Wheeler-DeWitt equation$^{[7]}$ 
\begin{equation}
\label{d-w}\frac{\partial ^2\psi }{\partial a^2}-\frac{6}{\kappa ^2a^2}\frac{%
\partial ^2\psi }{\partial \phi ^2}-\frac{144\pi ^4}{\kappa ^4}(K_ca^2-
\frac{\kappa ^2}3a^4V(\phi ))\psi =0,
\end{equation}
where $\psi $ is the wave function of the universe and $V(\phi )$ is a
potential of this scalar field. $K_c$ is the sign of the curvature term.
This is intricate equation, and it is hard to explain the meaning of the
wave function. In order to obtain a meaningful result a simplifying method
is needed. We adopt a new so-called ''twice loose shoe'' method$^{[10]}$,
i.e., to fix a variable and to let another variable vary respectively. When
one fixes at first step the scalar field $\phi $, i.e., $\partial \psi
/\partial \phi =0$, one has the equation on $a$ from Eq.(\ref{d-w}) 
\begin{equation}
\label{a-eq}\frac{\partial ^2\psi }{\partial a^2}=\frac{144\pi ^4}{\kappa ^4%
}(K_ca^2-\frac{\kappa ^2}3a^4V(\phi ))\psi .
\end{equation}
This is a following standard problem of non-relativity quantum mechanics,

$$
H\psi =E\psi ,\qquad H=-\frac{\hbar ^2}{2m}\frac{\partial ^2}{\partial x^2}%
+U(x),\qquad \frac{\partial ^2\psi }{\partial x^2}=\frac{2m}{\hbar ^2}%
(U(x)-E)\psi . 
$$
We know that Eq.(\ref{a-eq}) is a tunneling problem. Which potential and
total energy are

$$
U(a)=\frac{144\pi ^4}{\kappa ^4}(K_ca^2-\frac{\kappa ^2}3a^4V(\phi
)),\qquad E_a=0. 
$$
We note that only for $K_c=1$ the potential can form a barrier and has a
quantum tunneling effect. In this case the universe created is closed one.
In other hand, it is very hard to imagine that a flat or open universe with
real infinite volume can be created immediately from nothing by means of any
process. Therefore we suppose that only a finite closed universe can be
produced by quantum tunneling effect. $U(a)$ is a potential barrier, a
''particle'' will tunnel through it, and the tunneling probability is a
standard result$^{[6]}$ 
\begin{equation}
\label{tunnel}\rho _a=c_1\exp (-\frac{3\pi }{G\Lambda })=c_1\exp (-\frac{%
24\pi ^2}{\kappa ^4V(\phi )}).
\end{equation}
Note that it is independent on $a$. Here we don't debate the ''sign
problem'', we think this sign is more reasonable.

Serious problem is that there exists a so-called ''large field difficulty''
for this probability formula which statement is that if the potential is a
monotonous rising unbounded one, such as $m^2\phi ^2/2$, then the larger the 
$V$ is, the higher the probability is. Therewith the initial field value $%
\phi _0$ maybe tend towards infinity (at least possible up to value $\phi _q$
mentioned above), and the primary universe perhaps stays in state of the
excessive inflation. The key is that the most possible $\phi $ value is not
adjustable, which takes always its the largest one as possible as. In order
to obtain a suitable eternal chaotic inflation we have to choose a special
potential (undulate one?), or to cut off the applicable range of the
tunneling probability formula in case of unbounded potentials. Otherwise we
can not avoid this large field difficulty for this simple potential.

\section{The potential well effect of the scalar field}

{\hspace*{7mm}}After the tunneling, the $a$ takes transition from $a=0$ to a
non-zero value $a_0$ determined by the equation $U(a_0)=0$ of the turning
point, 
\begin{equation}
\label{phi-a0}\kappa ^2a_0^2V(\phi _0)=3,
\end{equation}
i.e., if we know $\phi _0$ of the primary universe, then we can calculate
its $a_0$.

When we use the loose shoe method at second step to fix the cosmic scale
factor $a$, i.e., $\partial \psi /\partial a=0$, we obtain the equation on $%
\phi $ from Eq.(\ref{d-w}) 
\begin{equation}
\label{phi-eq}\frac{\partial ^2\psi }{\partial \phi ^2}=8\pi ^4a^6(V(\phi
)-\frac{3}{a^2\kappa ^2})\psi .
\end{equation}
Comparing with the standard equation of quantum mechanics, it is a deep well
problem clearly, and its potential and total energy are 
$$
U(\phi )=8\pi ^4a^6V(\phi ),\qquad E_\phi =24\pi ^4\kappa ^{-2}a^4. 
$$
Remember that the $a$ is fixed here, therefore it doesn't obey a similar
equation $\kappa ^2a^2V(\phi )=3$ with Eq.(\ref{phi-a0}), otherwise the Eq.(%
\ref{phi-eq}) will be trivial. What value the $a$ can be fixed to? Let us do
an analysis. Before tunneling, nothing exits $a=0$, then $U(\phi )$ and $%
E_\phi $ all vanish, there is not any potential well for $\phi $. After
tunneling, the cosmic scale factor has been $a_0$ suddenly, in this time the
well is the most steep. In an actual process, the $a$ undergoes a tunneling,
we can imagine that $a$ changes from $0$ to $a_0$ in the fictive process of
the tunneling, therefore it is reasonable to suppose that $a$ takes a middle
value in the classical forbidden range between $0$ and $a_0$, i.e., $a=qa_0$%
, and an important parameter $q$ is introduced here which is a number factor
and is less than $1$.

When the potential is a square power one, i.e., the mass term $V=\frac
12m^2\phi ^2$, this is a harmonic oscillation problem and the Eq.(\ref
{phi-eq}) becomes as 
\begin{equation}
\label{harm}\frac{\partial ^2\psi }{\partial \phi ^2}=(4\pi
^4q^6a_0^6m^2\phi ^2-\frac{24\pi ^4}{\kappa ^2}q^4a_0^4)\psi . 
\end{equation}
We can do a variable transformation $\phi =sy$ to obtain the standard form 
\begin{equation}
\label{lam-n}\frac{\partial ^2\psi }{\partial y^2}=(y^2-\lambda )\psi
,\qquad s^{-2}=2\pi ^2q^3a_0^3m,\qquad \lambda =2n+1=\frac{12\pi ^2qa_0}{%
\kappa ^2m}. 
\end{equation}
The Eq.(\ref{harm}) has the solution

\begin{equation}
\label{phi-wave}\psi _\phi =c_2\exp (-\pi ^2q^3a_0^3m\phi ^2)\cdot w_n(\phi
), 
\end{equation}
where $w_n(\phi )$ is a polynomial for which the leading power term is $\phi
^n$. We shall see that only this highest term is important for our later
analysis since we want merely to consider a case of the larger values of
field $\phi $ or $y$ and power $n$.

Let us now consider the both effects comes from the tunneling and the well.
If unexpected complication is not involved, one can envisage that a combined
probability should be 
\begin{equation}
\label{tot-prob}\rho =\psi _\phi ^2\rho _a=c_1c_2^2\exp \left( 2n\ln (\kappa
\phi )-2\pi ^2q^3a_0^3m\phi ^2-\frac{48\pi ^2}{\kappa ^4m^2\phi ^2}\right)
\propto \exp (F(\phi )). 
\end{equation}
We see that just the well effect supplies a powerful suppression of this
probability for large $\phi $ value. This is what we hope in order to solve
the large field difficulty. The largest probability happens at the point $%
F^{\prime }(\phi _0)=0$, i.e., 
\begin{equation}
\label{max-prob}\frac{2n}{\phi _0}-4\pi ^2q^3a_0^3m\phi _0+\frac{96\pi ^2}{%
\kappa ^4m^2\phi _0^3}=0. 
\end{equation}
In this time we can substitute with $n=6\pi ^2q\kappa ^{-2}m^{-1}a_0$ and $%
a_0=\sqrt{6}\kappa ^{-1}m^{-1}\phi _0^{-1}$ in according to Eq.(\ref{lam-n})
and Eq.(\ref{phi-a0}) respectively, and obtain finally 
\begin{equation}
\label{phi-beg}\phi _0=4\sqrt{2/3}\cdot q^{-1}(2q^2-1)^{-1}\kappa ^{-1}, 
\end{equation}
i.e., the most possible inflaton field value when the universe is born in
quantum manner. We see that if the parameter $q$ is near to $2^{-1/2}$, the $%
\phi _0$ will approach to infinity. Of course it is not able to be taken
seriously due to approximation of our method. How to solve and understand
truly Eq.(\ref{d-w}) is a serious challenge!

\section{Startup, persistent and end of inflation}

{\hspace*{7mm}The transformation for the universe from quantum to classical
needs a process. For simpleness we assume that after} the universe is born
by quantum manner, it comes soon into the classical evolution. However the
universe does not come into inflation phase immediately due to a huge
curvature term. The equation of its motion is$^{[10]}$ 
\begin{equation}
\label{phi-a-mot}3\kappa ^{-2}(\frac{\dot a}a)^2=\frac 12\dot \phi ^2+\frac
12m^2\phi ^2-\frac{3\kappa ^{-2}}{a^2},\qquad \ddot \phi +3\frac{\dot a}%
a\dot \phi +m^2\phi =0, 
\end{equation}
with the initial condition $\phi _0$ of Eq.(\ref{phi-beg}), $\dot \phi _0=0$
and $a_0=\sqrt{6}\kappa ^{-1}m^{-1}\phi _0^{-1}$ at the starting time $t=0$
chosen by us. In this time the Hubble constant $H|_{t=0}$ is zero and
density ratio $\Omega |_{t=0}\rightarrow \infty $. At first we research how
the universe begins to inflate. The startup of the inflation in this case is
similar at all with the pure cosmological constant case $\Lambda =\kappa
^2V_\Lambda =3a_0^{-2}$. In the latter case, we have $H={\dot a}/a
=a_0^{-1}\tanh(a_0^{-1}t)$, 
but this expression only can applicable in the first few $a_0$%
's of time. After this time, i.e., about $t\geq 3a_0$, the Hubble constant
has arrived its highest value $H_b$ (index $b$ means beginning of inflation)
, and then $H$ begins to decrease its value, and the universe comes into the
normal slow rolling down and persistent inflation. This startup period of
about $t_b\simeq 3a_0$ is important to smooth whole primary universe and
maybe affect the final observed universe if it is in a gentle inflation.
Since $t_b$ is small, the difference between the original values $\phi _0$
of the inflaton and the values $\phi _b$, at which $H\approx H_b$ and the
real inflation begins, is small. We can use the approximation $\phi
_b\approx \phi _0$ to estimate the inflating e-folds. Our data simulation
supported these elementary analyses. The details of this model are very rich
which will be studied in our later work.

Before the parameter $q$ can be calculated exactly, we see that the most
possible initial field value is adjustable if we consider both effects of
tunneling and well. If $q$ is taken as $0.71\sim 0.72$, the $\phi _0$ is
near the $\phi _c$, the primary universe will evolve to an eternal chaotic
inflating universe, and the primary born probability lost its meaning as
explained by Guth$^{[8]}$. If $q$ is taken as $0.8$, then $\phi _0\simeq
15\kappa ^{-1}$ and $a_0\simeq 400\kappa $, the primary universe undergoes
only the gentle inflation, which inflating e-fold is about$^{[9]}$ 
\begin{equation}
\label{fold}N=\int M_p^{-2}VV^{\prime -1}d\phi =\kappa ^2(\phi _b^2-\phi
_e^2)/4\simeq \epsilon _b^{-1}/2=53.
\end{equation}
The concept of primary born probability is available in the gentle
inflation. At the ending of inflation, the cosmic scale factor is enough
large,%
$$
a_e=e^{53}a_b=10^{23}a_b, 
$$
however it is still a finite closed universe, although very flat. When the
inflaton rolls down to $\phi _e$ value, the inflaton field begins its fast
oscillation, the matter particles will be produced in succession, and later
on the universe comes into the radiation dominated period of the standard
big bang.

\section{More discussions about parameter $q$}

{\hspace*{7mm}}The parameter $q$ is very important in our model. Due to the
tunneling effect, $q$ should be less than $1$. However if $q$ is one
naively, we only have a little inflation, i.e., $\phi _0=3.3\kappa ^{-1}$
and $N\simeq 2$. In this case we can only put our hope on the small
probability events of the universe creation. However the relative
probability falls very rapidly in according to Eq.(\ref{tot-prob}). This is
not a way out.

We need to note that the role of the polynomial $w_n(\phi )$ which has an
obvious affect on $\phi _0$. However we can know the following point that
even if the largest contribution term is not the highest power term, the
modification is not too remarkable. The coefficient in the exponential of
the tunneling probability, in Eq.(\ref{tunnel}), such as $24$ or $12$, is
also not important, but its sign is important. The main affect comes from
the parameter $q$.

We can consider other potential such as $V(\phi )=\lambda \phi ^4$, however
we don't know its exact energy spectrum and are not able to estimate the
effect of the polynomial in the wave function. In the same time, there is
still a same problem whether the parameter $q$ has to be put in it to
reflect the tunneling character.

The parameter $q$ should be able to be calculated and only is dependent on
the unique model parameter, mass, if potential $m^2\phi ^2/2$ is used. Let
us imagine a wonderful prospect that if we know the function $q(m)$, then we
know $\phi _0(q(m))$. Moreover if the most possible field value $\phi _0$
was just the critical value $\phi _c(m)$ of the chaotic inflation, then we
could solve the unique parameter $m$ in the model. If this mass was just one
come from the CMBR, then it is an unbelievable high predicting ability for
this model!

Our idea may add some important constraint on the build of the inflation
models. It is not true that any potentials which can produce inflation can
also suitable to the quantum birth of the universe.

Anyhow the real meaning of the total Wheeler-DeWitt equation of Eq.(\ref{d-w}%
) is abstruse. We face a great challenge and opportunity.

\section{Conclusion}

{\hspace*{7mm}The universe born probability in the quantum cosmology has its
shortcoming if we only consider the tunneling effect, since the largest
probability happens at a place with the largest potential. For an unbounded
potential this is a disaster. The large potential has a large field value,
then the universe may stay in an excessive inflating state. It is necessary
to suppress the probability from the large field limit. This suppression
comes just from the potential well effect which is unconsidered previously.
In order to reflect the tunneling character a parameter }$q$ has to be
introduced, which should be able to be calculated from basic theory and only
is dependent on mass parameter if a simple potential $m^2\phi ^2/2$ is used.
Which state, i.e., gentle, chaotic or excessive inflation, the universe
stays in is subtly dependent on the parameter $q$. The primary universe
which has zero Hubble constant has to undergo a startup period to establish
its inflation, it can become smooth by means of this startup.

It is a pity that the ability to calculate the parameter $q$ out is absent
in the present. In fact many things are not known by us. We don't know
whether the ''twice loose shoe'' method is a proper one. We have to remember
that all of methods here used are only phenomenal. We don't know why only
four dimensions begins its expansion but why the extra-dimensions of the
superstring or membrane without expansion. There is some speculation in our
toy model. To enlighten more excellent new ideas or methods is what this
paper wants to pursue.

\bigskip

{\bf Acknowledgment:}

{\hspace*{5mm} This work is supported by The foundation of National Nature
Science of China, No.19777103. The author would like to thank useful
discussions with Profs. J.R.Bond, L.Kofman, U.-L.Pen during my visiting
Toronto University, and with X.-M.Zhang Y.-Z.Zhang and Y.-Q.Yu in China. }

\bigskip

{\bf References:}

[1] See e.g., P.J.E.Peebles, {\sl Physical Cosmology}, Princeton University
Press,

Prinston, 1993.

[2] A.H.Guth, Phys.Rev.D23(1981)347.

[3] A.Vilenkin, Phys.Rev.D37(1988)888.

[4] A.Linde, {\sl Particle Physics and Inflationary Cosmology},

1990 by Harwood Academic Publishers.

[5] S.W.Hawking and N.G.Turok, Phys.Lett.B425(1998)25.

[6] A.Linde, ''Quantum Creation of an Open Inflationary Universe'',

gr-qc/9802038.

[7] B.S.DeWitt, Phys.Rev.160(1967)1113;

J.A.Wheeler, in: Relativity, Groups and Topology, eds. C.M.DeWitt and

J.A.Wheeler (Benjamin, New York, 1968).

[8] A.H.Guth, ''Inflation and Eternal Inflation'', astro-ph/0002156.

[9] D.H.Lyth and A.Riotto, ''Particle Physics Models of Inflation and the

Cosmological Density Perturbation'', hep-ph/9807278.

[10] E.W. Kolb and M.S. Turner, {\sl The Early Universe}, Addison Wesley,

1990.

\end{document}